\documentclass[3p,times,twocolumn]{elsarticle}
\usepackage{ecrc}
\pdfoutput=1

\volume{00}

\firstpage{1}

\journalname{Nuclear Physics B Proceedings Supplement}

\runauth{T.~Th\"ummler et al.}

\jid{nuphbp}

\jnltitlelogo{Nuclear Physics B Proceedings Supplement}

\usepackage{upgreek}
\usepackage{graphicx}

\newcommand{\etal}{{\it et al.}\ }

\begin{document}

\begin{frontmatter}

\dochead{}

\title{Introduction to direct neutrino mass measurements and KATRIN}

\author[KIT]{T.~Th\"ummler}
\author[]{For the KATRIN Collaboration}

\address[KIT]{Institut f\"ur Kernphysik,
	Karlsruher Institut f\"ur Technologie (KIT), 
	Hermann-von-Helmholtz-Platz 1,
	76344 Eggenstein-Leopoldshafen,
	Germany}
	
\begin{abstract}
The properties of neutrinos and especially their rest mass play an important role at the intersections of cosmology, particle physics and astroparticle physics.
At present there are two complementary approaches to address this topic in laboratory experiments.
The search for neutrinoless double beta decay probes whether neutrinos are Majorana particles and determines an effective neutrino mass value.
On the other hand experiments such as MARE, KATRIN and the recently proposed Project 8 will investigate the spectral shape of $\upbeta$-decay electrons close to their kinematic endpoint in order to determine the neutrino rest mass with a model-independent method.
Here, because of neutrino flavour mixing, the neutrino mass appears as an average of all neutrino mass eigenstates contributing to the electron neutrino.
The KArlsruhe TRItium Neutrino experiment (KATRIN) is currently the experiment in the most advanced status of commissioning.
It combines an ultra-luminous molecular windowless gaseous tritium source with an integrating high-resolution spectrometer of MAC-E filter type.
It will investigate the neutrino rest mass with $0.2~\rm{eV/c^2}$ (90\% C.L.) sensitivity and allow  $\upbeta$ spectroscopy close to the $\rm{T_2}$ endpoint at 18.6 keV with unprecedented precision.
\end{abstract}


\end{frontmatter}

\section{Introduction}
Ever since Pauli postulated the Neutrino, its influence and significance for particle physics as well as for astrophysics has steadily been growing.
The observation of flavour oscillations of atmospheric and solar neutrinos, as well as of reactor and accelerator neutrinos at long baseline, is a compelling evidence that neutrinos are massive.
Due to this fact and owing to their large abundance in the universe, neutrinos are considered as the primary candidate for hot dark matter in cosmology.
Thus, they could play an important role for the evolution of large scale structures (LSS) in the universe.
On the other hand, the on-going investigations of neutrino properties, their mass hierarchy and especially their rest mass will open a door to the understanding of the origin of mass.
Here, a precision measurement of the neutrino rest mass can discriminate between different $\upnu$-mass models, in particular whether they are of hierarchical type $(m_1 \ll m_2 \ll m_3)$ or of quasi-degenerate type $(m_1 \simeq m_2 \simeq m_3)$. Although experiments on neutrino flavour oscillation provide compelling evidence that neutrinos are massive, they cannot provide an absolute mass value.
Therefore, other methods have to be applied.

Cosmological investigations provide very sensitive methods to determine an estimate for the neutrino mass.
Here, the sum of all neutrino mass eigenstates is being measured, since in cosmological processes the flavour cannot be distinguished.
Using only WMAP7 data, an upper limit of 1.2~eV$/c^2$ is determined for the sum of neutrino mass eigenstates \cite{Han2010}.
Upcoming data from the Planck satellite as well as from LSST and weak lensing surveys have the potential of probing the sum of neutrino masses with 50 meV$/c^2$ sensitivity.
However, while cosmologcal studies offer a very sensitive approach, one has to emphasize its generic model dependence, which easily leads to a factor of 2 of uncertainty in the neutrino mass prediction \cite{GonGar2010}.

Therefore, it is essential to probe the neutrino mass with sub-eV sensitivity in laboratory experiments.

\section{Direct $\upnu$-mass measurements in laboratory experiments}

In general, there are two approaches which are complementary in their physics objectives.
One is the search for neutrinoless double $\upbeta$-decay ($0\upnu\upbeta\upbeta$), the other is the precise spectroscopy of $\upbeta$-decay at its kinematic end-point.

The $0\upnu\upbeta\upbeta$ process requires the neutrino to be a Majorana particle, thus it appears identical to its anti-particle.
From neutrinoless double $\upbeta$-decay an effective Majorana mass  $m_{\beta\beta}$ is determined, which corresponds to the coherent sum of all mass eigenstates $m_{\nu_{\rm i}}$ with respect to the PMNS mixing matrix $U_{\rm ei}$.
\begin{equation}
\label{doppelbeta}
m_{\beta \beta} =  \left |\sum_{\rm i} U_{\rm ei}^2 \cdot m_{\nu_{\rm i}}\right |,
\end{equation}
Here, not $\left | U_{\rm ei} \right | ^2$ but $U_{\rm ei}^2$ appears, hence $m_{\beta \beta}$ depends on complex CP-phases with the possibility of cancellations.
This generic fact has to be taken into account in terms of a model-dependence when comparing claims or limits like $m_{\beta\beta} < 0.35~{\rm eV}/c^2$ \cite{Klapdor2001} to results from single $\upbeta$-decay experiments.
Nevertheless, upcoming $0\upnu\upbeta\upbeta$ experiments like GERDA, MAJORANA, SNO+, EXO, and CUORE have the potential to probe $m_{\beta\beta}$ in the 20 - 50 meV/$c^2$ region.
In \cite{Rodejohann_Nu2010, Pavan_Nu2010, Dolinski_Nu2010, Nakamura_Nu2010}, and \cite{Simkovic_Nu2010} the experimental and theoretical status of this field as well as the influences by matrix elements are discussed in more detail.

Experiments investigating single $\upbeta$-decay offer a direct and model-independent method to determine the absolute neutrino mass, since they rely only on the relativistic energy-momentum relation and energy conservation \cite{Valle_Weinheimer, Otten_Weinheimer2008}.
In single $\upbeta$-decay the squared neutrino mass $m_{\nu_{\rm e}}^2$ is determined as an incoherent sum -- in contrast to (\ref{doppelbeta}) -- of all mass eigenstates according to the PMNS matrix.
\begin{equation}
\label{singlebeta}
m_{\nu_{\rm e}}^2 = \sum_{\rm i} \left | U_{\rm ei} \right | ^2 \cdot m_{\nu_{\rm i}}^2
\end{equation}
Currently, the best experimental bounds from single $\upbeta$-decay ($m_{\nu_{\rm e}} < 2.3~{\rm eV}/c^2$) have been determined in the tritium $\upbeta$-decay experiments in Mainz \cite{Kraus2005} and Troitzk \cite{Lobashev2003}, which have reached their sensitivity limit.
At present new experiments are being assembled or designed which will increase the experimental precision by two orders of magnitude, thus increasing the sensitivity on $m_{\nu_{\rm e}}$ by one order of magnitude to 200~meV/c$^2$.

The basic principle applied in this model-independent method is based on kinematics and energy conservation only, with the squared neutrino mass eigenstate $m_{\nu_{\rm i}}^2$ appearing in the phase space factor.
For a detailed description including final states and relativistic kinematics refer to \cite{Valle_Weinheimer, Otten_Weinheimer2008}.
\begin{eqnarray}
\nonumber
\frac{{\rm d}{\Gamma_{\rm i}}}{{\rm d}E} &=& \frac{{\rm G}_{\rm F}^2}{2 \pi^3 \hbar^7 c^5} \cos^2(\theta_C)|\mathcal{M}|^2 F(Z+1,E)\cdot\\
& & p \cdot (E+m_e c^2) \cdot (E_0-E) \cdot\\
\nonumber
& &\sqrt{(E_0-E)^2-m_{\rm i}^2 c^4}  \,\, \Theta(E_0-E-m_{\rm i}c^2),
\label{equ:spek1}
\end{eqnarray}
Since the experimental resolution at present does not allow to resolve individual mass eigenstates, only a combined spectrum can be recorded, according to
\begin{equation}
\frac{{\rm d}\Gamma}{{\rm d}E} = \sum_{i = 1}^{3}\frac{{\rm d}{\Gamma_{\rm i}}}{{\rm d}E}.
\end{equation}
The electron neutrino mass parameter corresponding to (\ref{singlebeta}) is derived by analyzing the spectrum close to the end-point, where $(E_0 -E)$ is small and the mass term $m_{\rm i}$ becomes significant.
A non-zero neutrino mass will not only shift the end-point, but also change the spectral shape.
The count rate at the end-point is very low and the absolute Q-value is not known precisely enough, therefore it is far more important to measure the shape of the spectrum, rather than the end-point shift.
A low end-point $\upbeta$ source for a large fraction of electrons in the end-point region is a key requirement as well as a high source luminosity and high energy resolution in addition to very low background.
For single $\upbeta$-decay investigations there are two complementary approaches with different systematics.
\begin{enumerate}
\item The calorimeter approach, where the source is identical to the detector.\\
Here, metallic or dielectric $^{187}$Re crystal bolometers are used to capture the entire $\upbeta$-decay energy as differential energy spectrum.
With $2.47$~keV $^{187}$Re has the lowest end-point, but due to its rather long half-life of $4.3 \times 10^{10}$~y the activity is low.
Since bolometers are modular, their number can be scaled in order to increase the sensitivity.
This approach is being followed in the MARE experiment.\\
\item The spectrometer approach, where an external tritium source is used.\\
In this case, the kinetic energy of the $\upbeta$-electrons is analyzed as integral spectrum by an electrostatic spectrometer.
Here, the source material is high purity molecular tritium $\rm T_2$ with a low end-point at $18.6$~keV and a short half-life of $12.3~y$, providing a high activity.
This approach implies strict scaling laws for the size of the spectrometer and reaches its ultimate size and precision in the KATRIN experiment.
\end{enumerate}
Besides these approaches new ideas have recently come up.
The Project 8 proposal \cite{Monreal2009} aims to make use of radio-frequency spectroscopy in order to measure the kinetic energy of $\upbeta$ electrons from a gaseous $\rm T_2$ source.
Here, an array of antennas would capture the coherent cyclotron radiation emitted by the $\upbeta$-decay electrons when moving through a homogenous magnetic field.
In \cite{Monreal2009} the authors estimate an ultimate sensitivity of $0.1$~eV for this method.
Currently, a proof-of-principle experiment is in preparation in order to show that it is feasible to detect electrons and to determine their kinetic energy with this method.
For details please see \cite{Formaggio_Nu2010}.

\begin{figure*}
\begin{center}
\includegraphics*[width=0.98\textwidth]{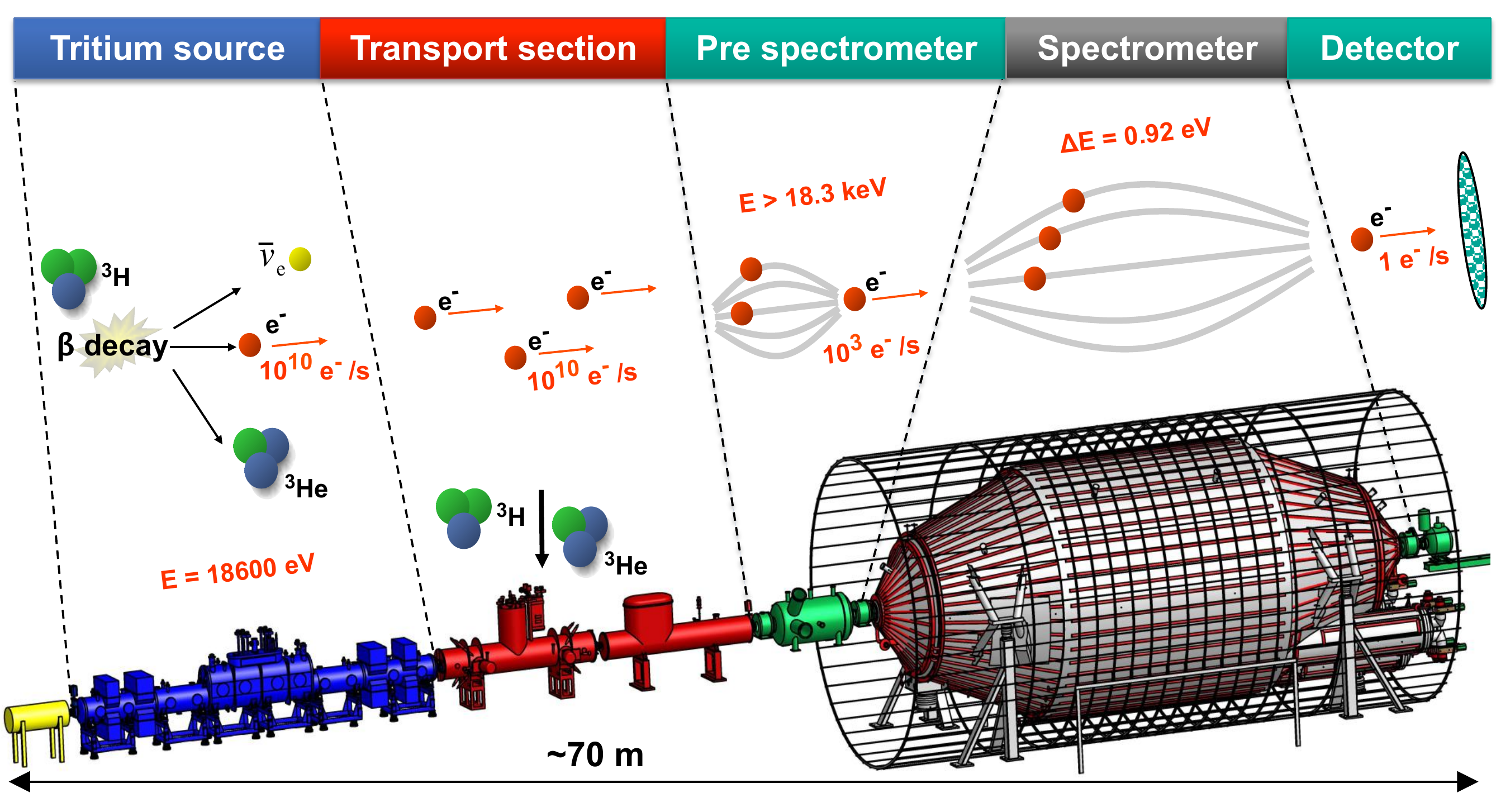}
\end{center}
\caption{Overview of the KATRIN main beam-line.
High purity $\rm T_2$ gas is being injected into the source tube.
The electrons from $\upbeta$-decay leave the source and are guided by magnetic fields through the transport section, while the remaining gas is being removed by active and cryogenic pumping.
The pre-spectrometer filters out the low-energy part of the spectrum, thus only electrons close to the endpoint region can enter the main spectrometer for the precise energy analysis.
Transmitted electrons are then detected by a low background Si-PIN detector system.
}
\label{fig:KATRIN}
\end{figure*}

\section{The MARE experiment}

As outlined above, MARE (Microcalorimeter Arrays for a Rhenium Experiment) uses $^{187}$Re as $\upbeta$ emitter.
The long half-life of this isotope is the result of the first order unique forbidden transition $^{187}_{85}{\rm Re} \rightarrow  {^{187}_{86}{\rm Os}} + e^- + \bar \nu_e$.
MARE will thus be performed in two phases with increased sensitivity.
It is the successor of the MANU and MIBETA experiments, which derived an upper limit of $m_\nu < 15$~eV from $10^6$ $\upbeta$-decays.

Phase I of MARE aims to improve the current sensitivity on $m_\nu$ by one order of magnitude by increasing the statistics to $\approx 10^{10}$ $\upbeta$-decays.
This will result in a sensitivity comparable to the Mainz and Troitzk experiments ($m_\nu \approx 2$~eV), thus scrutinizing their upper limit.
Using calorimeter pixel arrays, it will take three years to reach the intended number of decays.
This time is used in parallel to finish the R\&D program in order to improve the sensor technology for the next phase II.

The goal of MARE phase II is to improve the sensitivity by another order of magnitude with increased statistics of $10^{14}$ $\upbeta$-decays.
With a sensitivity of $m_\nu \approx 0.2$~eV MARE will then be in a position to scrutinize a KATRIN result with a completely independent method and different systematics.
In order to reach the projected sensitivity the operation of 50000 bolometer pixels is required over a measurement time of at least 5 years.

Detailed information about the present status and the progress of both MARE phases can be found in \cite{Nucciotti_Nu2010}, in addition to a review of the investigation of $^{163}$Ho as alternative $\upbeta$ emitter. 

\section{The KATRIN experiment}

The KATRIN (KArlsruhe TRItium Neutrino) experiment \cite{KATRIN-LOI, KATRIN_design_report} makes use of a molecular gaseous tritium source and an electrostatic energy filter based on the MAC-E filter principle \cite{picard-mace}.
An overview of the main beam-line is given in figure \ref{fig:KATRIN}.
The MAC-E filter uses magnetic adiabatic collimation in combination with an electrostatic energy filter.
Electrons emitted in the source at high magnetic field enter the spectrometer, where the magnetic field drops by several orders of magnitude.
Maintaining a strictly adiabatic regime with full energy conservation, the magnetic gradient force transforms the cyclotron energy of the isotropically emitted electrons into the longitudinal component.
At the same time, the increasing potential of the electrostatic filter slows down the electrons by reducing their longitudinal energy.
Both effects have to be aligned very carefully for a proper energy analysis.
Only electrons which are able to cross the potential in the spectrometer are counted, thus the MAC-E filter acts as high-pass filter for electrons.
Its relative resolution is defined by the magnetic field ratio between minimum field in the center of the spectrometer and maximum field at the pinch magnet.
This defines the overall electromagnetic design of KATRIN and the size of its components.

KATRIN faces challenges in all key technologies which are applied in the experiment.
As an example, about $10^{11}$ electrons/s are emitted in the source, while at the far detector side a background rate of the order of $10^{-2}$~cps has to be reached.
The source column density, temperature and pressure has to be kept stable at the $10^{-3}$ level.
The transport section has to reduce the tritium gas flow by a huge amount of $10^{14}$, in order not to increase the background in the spectrometer section.
In addition the decay electrons have to be transported fully adiabatically on the meV-scale over a distance of about 50~m.
In the spectrometer region, carefully designed electromagnetic field conditions have to be implemented with high voltage stability on the $10^{-6}$ scale for operating voltages up to $35$~kV.
In particular, electromagnetic conditions leading to Penning-like trapping conditions have to be avoided in order to reach low background rates.
The following subsection addresses the details of some key components and their present status.

\subsection{The KATRIN setup}
The KATRIN setup stretches over 70~m.
The windowless gaseous tritium source (WGTS) provides a high activity of $10^{11}$ $\upbeta$-decays per second.
The tritium gas is injected in the center part of the source beam-tube and flows to both ends, where it is removed again by pumps.
Electrons emitted by the T$_2$-decay are guided by strong magnetic fields ($B = 3.6$~T in the source, $B = 5.6$~T in the transport section) to the transport section and finally to the spectrometer section, while at the same time retaining the gas flow by 14 orders of magnitude.
The pre-spectrometer -- which is a MAC-E filter operating at moderate resolution of about $100$~eV -- can be used to transmit only electrons with energies close to the T$_2$ end-point.
In this operation mode, only electrons of the end-point region would enter the main spectrometer for the precise energy analysis.
The main spectrometer is a MAC-E filter and offers a resolution of $0.93$~eV resolution for $18.6$~keV electrons by applying a magnetic field ratio of $1/20000$.
The latter defines the size of the main spectrometer, since the full magnetic flux tube of the source has to be analyzed.

\subsubsection{Windowless gaseous tritium source}
In order to meet its sensitivity goal, KATRIN requires a stable source with high luminosity, and low systematic effects.
The design luminosity of $1.7 \times 10^{11}$~Bq requires $5 \times 10^{19}$ T$_2$ molecules to be injected per second into the source tube, which adds up to a throughput of 40~g per day.
A unique research facility in Europe providing the corresponding tritium technology is Tritium Laboratory Karlsruhe (TLK), located at the KIT Campus North site.
The tritium purity has to be kept above 95\%, the source temperature at $T=27$~K, and the injection pressure at $10^{-3}$~mbar.
In order to achieve a stable column density all these parameters have to be stable on the $10^{-3}$ level.

At TLK a sophisticated $\rm T_2$ loop system has been commissioned for KATRIN, providing a stable T$_2$ injection rate and a high purity isotope separation stage \cite{Fischer_Nu2010}.
Being designed for a stability of $10^{-3}$, test runs showed that the loop system reaches a $10^{-4}$ stability level for over 4 months.

\begin{figure*}
\begin{center}
\includegraphics*[width=\textwidth]{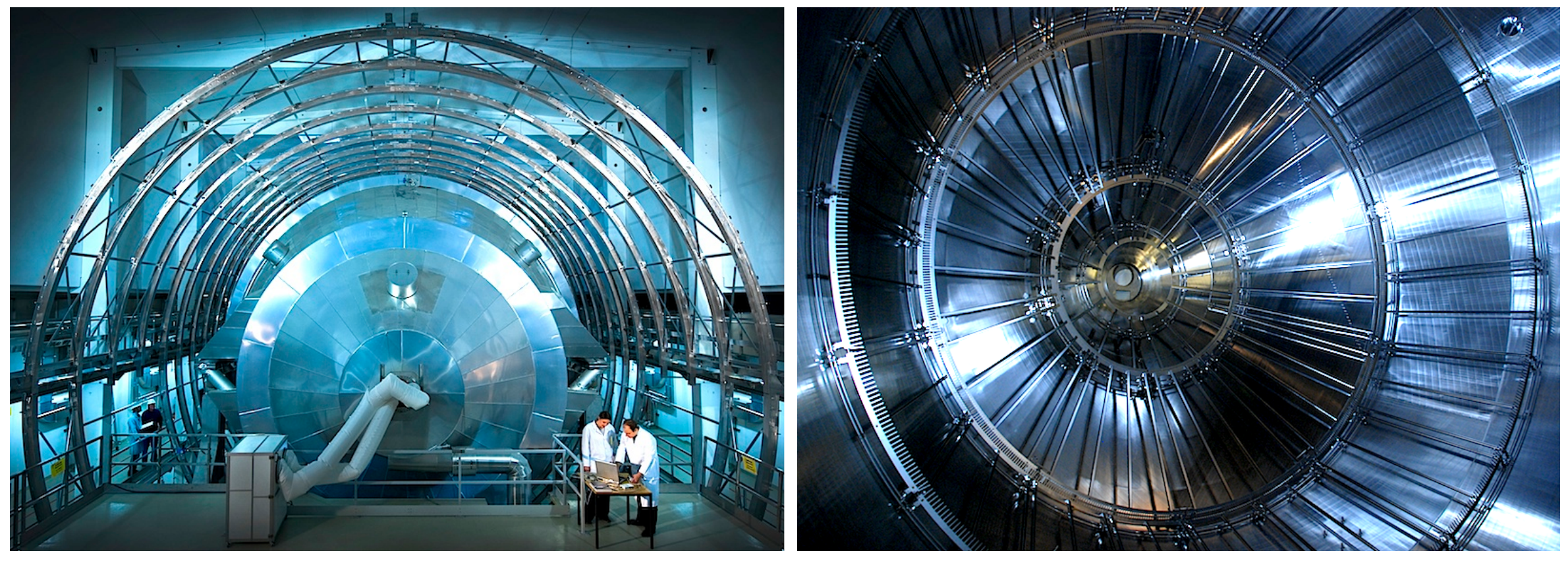}
\end{center}
\caption{
The left picture shows a view of the main spectrometer's vacuum vessel along the beam-axis (without pre-spectrometer installed) surrounded by the air-coil support structure.
The picture on the right shows a view from inside the main spectrometer during the installation of the wire electrode modules in the conical part of the vessel (wires are hardly visible on this scale).
}
\label{fig:mainspec}
\end{figure*}

A new concept of temperature stabilization is applied to the WGTS beam tube ($l = 10$~m, $d =  9$~cm)  in order to achieve a stability of $\Delta T \leq 30~{\rm mK}$ at the source operating temperature.
The inner beam-tube is directly coupled to a two phase Ne thermosiphon \cite{Grohmann2009}.
This novel concept is presently being tested at the WGTS demonstrator, which is the inner part of the WGTS cryostat without superconducting magnets. 
After the demonstrator tests the WGTS cryostat will be completed in order to be ready for implementation into the KATRIN beam-line in 2012.

\subsubsection{The transport section}
Adjacent to the WGTS the transport section guides the decay electrons to the spectrometer section by strong magnetic fields of $B = 5.6$~T.
Tritium retention by the factor $10^{14}$ is achieved in two steps.
First the differential pumping section DPS uses active pumping to provide a T$_2$ retention of $10^7$.
The DPS2-F cryostat is on-site and its commissioning has recently been completed.
The cryogenic pumping section CPS uses cryo-sorption to achieve a further retention of $> 10^7$.
Here, the liquid He cold inner beam-tube is covered with Ar frost, providing a large surface to cryo-sorp the T$_2$.
The CPS concept has been successfully tested in \cite{Oleg2009, Eichelhardt2009}, the cryostat is expected to be on-site in mid 2011.

\subsubsection{The pre-spectrometer test setup}
The spectrometer section consists of two MAC-E filters.
The pre-spectrometer is a MAC-E filter with sufficient resolution to cut the low energy part of the $\upbeta$ spectrum at about 300~eV below the end-point, if required.
At present it acts as an important test setup, where all major technologies for the main spectrometer have been tested, i.e. the vacuum concept, the electromagnetic design as well as electron transmission properties and the detector concept.
Especially the electromagnetic design will have a strong impact on the background reduction.
Even tiny cm$^3$-sized Penning-like traps close to the ground electrode can cause background rates of $> 10^3$~cps.
Based on the advanced KATRIN field calculation code, the electrode shape has been refined avoiding these traps and resulting in a background level of a few $10^{-3}$~cps.
This rate cannot be distinguished from the detector background, hence the pre-spectrometer is considered as quasi background free.
Remaining background features have originated from Rn decays in the volume, which come from auxiliary components and the SEAS non-evaporable getter material.
The auxiliary parts involved have been removed and Rn from the getter material has been eliminated by LN$_2$ baffles, for details see \cite{Fraenkle-phd}. 
The pre-spectrometer will be integrated into the beam-line in 2011.

\subsubsection{The main spectrometer}
The main spectrometer analyses the $\upbeta$-decay electrons of the end-point region with a resolution of $0.93$~eV and has to operate with a background rate at the $10^{-3}$~cps level.
Therefore, background from low-energy electrons created by cosmic muons in the vessel wall have to be rejected.
For this purpose not only a magnetic shielding is being relied upon, but also a UHV compatible wire electrode system (240 modules, 23000 wires, $0.2$~mm precision) is currently being installed, see figure \ref{fig:mainspec}.
Vacuum tank and wire electrode provide the precise retarding potential for the energy filter and are supplied by high-voltage equipment at a precision level of $10^{-6}$.
This level is of key importance for the $\upnu$-mass measurement, thus it will be monitored by ultra-precise high-voltage dividers \cite{Thuemmler2009, Marx2001} and a monitor spectrometer beam-line.
The latter uses the same retarding potential and measures mono-energetic electron sources.
Since the electromagnetic properties of the MAC-E filter are extremely sensitive to distortions of the magnetic field, an air coil system has been installed.
As shown in figure \ref{fig:mainspec} it consists of large coils around the vacuum vessel.
The air coil system compensates the earth magnetic field and in addition it is used for the fine tuning of the magnetic flux tube.

\subsubsection{The detector setup}
The transmitted electrons are counted by a detector based on a monolithic 148 pixel Si-PIN diode with an energy resolution of about 1~keV, and the ability to detect rates from $10^{-3}$~cps in the end-point region and up to $10^{3}$~cps during calibration runs, while keeping the background low.
Mechanically it is designed like a dartboard with 9~cm diameter and 12 concentric rings with $30^\circ$ segmentation.
The detector setup including magnets, vacuum components and DAQ is being delivered for system integration in early 2011 \cite{Wall_Nu2010, Renschler_Nu2010}.  

\section{KATRIN outlook and sensitivity}

After commissioning and testing, the complete system integration is planned for late 2012.
As soon as the setup is in measurement condition, KATRIN will take 3 years of data, which corresponds to 5 years real-time.
The statistical uncertainty is estimated to be $\Delta m_{\rm stat}^2 = 0.018$~eV$^2/c^4$.
Likewise, the systematic uncertainty is being restricted to $\Delta m_{\rm syst}^2 < 0.017$~eV$^2/c^4$ in order to minimize the total uncertainty to $\Delta m_{\rm tot}^2 = 0.025$~eV$^2/c^4$.
This leads to the KATRIN design sensitivity of $m_\upnu = 0.2$~eV$/c^2$ (90\% CL) and the discovery potential of $m_\upnu = 0.35$~eV$/c^2$ with 5$\upsigma$ significance.

\section{Summary}

The on-going efforts to measure the absolute mass of neutrinos is well motivated by recent results in particle and astroparticle physics.
The investigation of $\upbeta$-decay offers a model-independent method for determining the $\upnu$-mass.
The Re-based MARE experiment is going to check the present limits, while at the same time improving the detector technology.
In future, MARE will be able to scrutinize a KATRIN result, after operating 50000 bolometer pixels for at least 5 years.
New concepts for $\upbeta$-spectroscopy such as Projects 8 will open new ways to optain sub-eV sensitivities and should be matured in an R\&D phase.
The estimated sensitivity of the tritium-based KATRIN experiment is $m_\upnu = 0.2$~eV/$c^2$ (90\% CL).
Its construction is proceeding well with several of the major components being already on-site.
The main spectrometer test program will start in 2011 and the complete system integration is planned for 2012.


\section*{References}

\begin{thebibliography}{00}
\bibitem{Han2010} S.~Hannestad \etal, arXiv:1004.0695
\bibitem{GonGar2010} M.~C.~Gonzalez-Garcia \etal, JHEP {\bf 08} (2010) 117,  arXiv:1006.3795
\bibitem{Klapdor2001} H.~V.~Klapdor-Kleingrothaus \etal, Eur. Phys. J. A {\bf 12} (2001) 147
\bibitem{Rodejohann_Nu2010} W.~Rodejohann, "Neutrinoless Double Beta Decay in Particle Physics", these proceedings
\bibitem{Pavan_Nu2010} M.~Pavan, "Introduction to double beta decay experiments and CUORE", these proceedings
\bibitem{Dolinski_Nu2010} M.~Dolinski, "The Enriched Xenon Observatory (EXO)", these proceedings
\bibitem{Nakamura_Nu2010} K.~Nakamura, "Scintillator-based experiments in Double Beta Decay", these proceedings
\bibitem{Simkovic_Nu2010} F.~Simkovic, "Matrix Elements for Double Beta Decay", these proceedings
\bibitem{Valle_Weinheimer} S.~S.~Masood \etal, Phys. Rev. C {\bf 76} (2007) 045501
\bibitem{Otten_Weinheimer2008} E.~W. Otten and Ch. Weinheimer, Rep. Prog. Phys. {\bf 71} (2008) 086201
\bibitem{Kraus2005} C.~Kraus \etal, Eur. Phys. J. C. {\bf 40} (2005) 447 
\bibitem{Lobashev2003} V.~M. Lobashev, Nucl. Phys. {\bf A719} (2003) 153c 
\bibitem{Monreal2009}  B.~Monreal \etal, Phys. Rev. D {\bf80} (2009) 051301(R)
\bibitem{Formaggio_Nu2010} J.~A.~Formaggio, "Project 8: Measuring Neutrino Mass using Radio-Frequency Techniques", these proceedings
\bibitem{Nucciotti_Nu2010} A.~Nucciotti, "Neutrino mass calorimetric searches in the MARE experiment", these proceedings
\bibitem{KATRIN-LOI} A.~Osipowicz \etal, (KATRIN Collaboration), {\tt arXiv:hep-ex/0109033} (2001)
\bibitem{KATRIN_design_report} J.~Angrik \etal, FZK Scientific Report 7090, {\tt bibliothek.fzk.de/zb/berichte/FZKA7090.pdf}
\bibitem{picard-mace} A.~Picard \etal, Nucl. Instr. Meth. {\bf 63} (1992) 345
\bibitem{Fischer_Nu2010} S.~Fischer, "Laser Raman Spectroscopy for KATRIN", these proceedings
\bibitem{Grohmann2009} S.~Grohmann, Cryogenics, Volume 49, Issue 8 (2009) 413-420

\bibitem{Oleg2009} O.~Kazachenko \etal, Nuclear Instruments and Methods in Physics Research A {\bf 587} (2008) 136
\bibitem{Eichelhardt2009} F.~Eichelhardt \etal, Fusion Science and Technology {\bf 54} (2008) 615
\bibitem{Fraenkle-phd} F.~M.~Fr\"ankle, Dissertation (2010), {\tt digbib.ubka.uni\--karlsruhe.de/volltexte/1000019392}
\bibitem{Marx2001} R.~Marx, IEEE Trans. on Instr. and Meas. 50 No. 2 (2001) 426-429
\bibitem{Thuemmler2009} T.~Th\"ummler \etal, New J. Phys. {\bf 11} (2009) 103007
\bibitem{Wall_Nu2010} B.~Wall, "Overview of the KATRIN Detector Section", these proceedings
\bibitem{Renschler_Nu2010} P.~Renschler, "MC Simulations of the detector response to low energy electrons for KATRIN", these proceedings



\end{thebibliography}
\end{document}